\documentclass[12pt, fleqn]{article}
\usepackage[mathcal]{euscript}
\usepackage{graphicx}
\usepackage{amssymb}
\usepackage{epstopdf}
\usepackage{latexsym}
\usepackage{indentfirst}
\usepackage{marvosym}
\usepackage{hyperref}

\def\K3{\mathrm K3}
\def\CY#1{\mathrm{CY}_{#1}}
\def\SU#1{SU({#1})}
\def\U#1{U({#1})}
\def\SO#1{SO({#1})}
\def\O#1{O({#1})}

\def\double #1{#1{\hbox{\kern-2pt $#1$}}}

\begin{document}
\setcounter{page}0


\thispagestyle{empty}

~
\vspace{-70pt}

\begin{flushright}
\makebox[0pt][b]{YITP-SB-06-48}
\end{flushright}

\vspace{40pt}

\center{{\LARGE Covariant $N=2$ heterotic string in four dimensions}

\vspace{30pt}

{\large William D. Linch, {\sc iii}${}^{\small \mbox\Pisces}$\
and Brenno Carlini Vallilo${}^{\small \mbox \Libra}$ }
}

\vspace{10pt}

\center{
${}^{\small \mbox\Pisces}${\em C.N. Yang Institute for Theoretical Physics\\ and\\ Department of Mathematics} \\
{\it SUNY, Stony Brook, NY 11794-3840, USA}

${}^{\small \mbox\Libra}${\em Instituto de F\'{\i}sica, Universidade de S\~ao Paulo\\ C.P. 66318, 05315-970, S\~ao Paulo, SP, Brasil}
}
\vspace{10pt}

\abstract{
We construct a covariant formulation of the heterotic superstring on $\K3 \times T^2$ with manifest $N=2$ supersymmetry. We show how projective superspace appears naturally in the hybrid formulation giving a (partially) geometric interpretation of the harmonic parameter. The low-energy effective action for this theory is given by a non-standard form of $N=2$ supergravity which is intimately related to the $N=1$ old-minimal formulation.
This formalism can be used to derive new descriptions of interacting projective superspace field theories using Berkovits' open string field theory and the heterotic Berkovits-Okawa-Zwiebach construction.
}

\vspace{1.5in}
\begin{flushleft}
\makebox[0pt][t]{${}^{\small \mbox\Pisces}$ wdlinch3@math.sunysb.edu}\\
\makebox[0pt][t]{${}^{\small \mbox\Libra}$ vallilo@fma.if.usp.br}
\end{flushleft}

\newpage


\section{Introduction}

In this note we present a new formulation of the heterotic
string in four dimensions. This formulation can be used to describe
compactifications with $N=1$ supersymmetry which do not come from a
$\CY3$ compactification and thus cannot be obtained by the usual hybrid method. The formalism extends the known superspace
covariant quantization approaches to string theory \cite{covariant}
to include the $N=2$ heterotic string on backgrounds of the form $\K3
\times T^2$.

Covariant formulations of superstring theories, by definition, depend
heavily on our understanding of the structure of off-shell
superspaces. Over the past two decades the latter have been
investigated intensively with varying degrees of success. The main
problem in this field consists of finding a formulation in which the
basic constraints defining the superfield representations can be
solved in terms of unconstrained superfields (prepotientials). We
have come to learn that this requires the introduction of an infinite
number of auxiliary fields when the number of (real) supercharges
exceeds 4. These auxiliary fields are naturally organized in
``harmonics''; they come from an expansion in a parameter
which describes a coordinate on an auxiliary space related to the
$R$-symmetry of the extended theory. A concrete realization of this
is the 4-dimensional, $N=2$ harmonic superspace of Galperin, Ivanov,
Kalitzin, Ogievetski, and Sokachev \cite{Galperin:1984av} in which
the auxiliary parameter is a zwei-bein on the $\SU2/\U1=S^2$ coset
where the $\SU2$ is the $N=2$ $R$-symmetry group. A second such
realization is the projective superspace of Karlhede, Lindstr\"om,
and Ro$\check{\rm c}$ek \cite{Karlhede:1984vr} in which the auxiliary
parameter is a holomorphic coordinate on the punctured complex plane
$\mathbb C^*$. Although this formulation was constructed
independently of the harmonic superspace, it was shown explicitly by
Kuzenko how the former is realized as a singular limit of the latter
\cite{Kuzenko:1998xm}. This ``double puncture'' limit has a natural
interpretation as dimensional reduction in superspace \cite{Kuzenko:2005sz} if
both formalisms are extended to 5 or 6 dimensions in the sense that
the projective superspace necessarily breaks the Lorentz invariance
to $\SO{3,1}$.

In the hybrid formalism for the $\K3\times
T^2$ compactification of the heterotic $N=(2,0)$ description of the
superstring we propose here, it is useful to consider the compactification as the
heterotic model on a $\K3$ surface $S$ with $N=(1,0)$ supersymmetry
in the 6-dimensional target further reduced on a torus
$T^2$. The resulting theory has $N=2$ supersymmetry in the
4-dimensional target $M$. The natural superspace for describing
6-dimensional theories with $N=(1,0)$ supersymmetry is the original
harmonic superspace \cite{Galperin:1984av} extended by two dimensions.
Further reduction on $T^2$ puts the theory, through the
double puncture limit, in the projective superspace.\footnote{The
4-dimensional, $N=2$ harmonic superspace was extended to 5 dimensions
and reduced to projective superspace in detail in \cite{Kuzenko:2005sz}. Aspects
of the extension of the projective formalism to 6 dimensions were
investigated in \cite{ProjExt}.} Therefore, from a geometrical point
of view, it is natural that the heterotic description on $\K3\times
T^2$ lives in projective superspace. In fact, we will find that
similarly to the 6-dimensional type II description on $\K3$
constructed by Berkovits \cite{Berkovits:1999du}, the hybrid will in
our case also provide a natural candidate for the projective
parameter. Contrary to that case, however, our projective parameter will have a
(partially) geometric interpretation, being related to the RNS fermion of the
torus $T^2$.

Perhaps it is prudent to emphasize that, although the formalism will
display many similarities with the construction given by Berkovits in
\cite{Berkovits:1999du}, the present construction is not directly
related to the latter by compactification. This will become clear below, 
where we will see that the projective constraint
introduced here cannot be obtained from that in \cite{Berkovits:1999du}.
Indeed, the two formalisms should only be related insofar as the
6-dimensional formalism reduces to the standard hybrid after a
further compactification breaking half of the 16 supercharges. The
relation should then follow from string-string
duality mapping the heterotic description on $\K3\times T^2$ to the
type II description on an $\K3$-fibered Calabi-Yau.

This note is organized as follows. In section
\ref{Sec:HybridFormalism} we introduce the conformal field theory for
the heterotic string compactified on $\K3\times T^2$ and discuss its
worldsheet symmetries and constraints. Section \ref{Sec:VtexOps} is
dedicated to deriving the physical vertex operators from these
constraints, of which there are two types. We consider the two cases 
in turn in sub-sections \ref{Section:CIVtex} and \ref{Section:Supermodulons}. 
We conclude this section with a summary of the massless spectrum in 
projective superspace. 
In section \ref{Section:SG}, we find the explicit realization by the hybrid
formalism of Siegel's proposal for the conformal supergravity
prepotential of heterotic $N=2$ superspace supergravity
\cite{Siegel:1995px} and reproduce the super-dilaton
(compensator) structure of this theory. Due to Siegel's work, 
we can immediately write down the gravitational
and gauge
part of the effective action as it follows from the anomaly
cancellation mechanism.
Similarly, the gravitational couplings to the supermoduli follow from
superspace rules. 
We conclude in section \ref{Sec:Outlook} with examples of applications of our
results, an interesting one of which constitutes the first closed string field theory
with $N=2$ target space supersymmetry.

\section{Hybrid formalism on $\K3\times T^2$}
\label{Sec:HybridFormalism}
In this section we propose a hybrid formalism for the $\K3\times T^2$
compactification of the heterotic $N=(2,0)$ description of the
superstring. The action of the heterotic string with $N=(2,0)$
worldsheet supersymmetry with a target space $M\times \K3\times T^2$
is given by
\begin{eqnarray}
\label{lagrangian}
S=\int d^2z \Big[ \partial x_m \bar \partial x^m
+p_{\alpha} \bar \partial \theta^{ \alpha}+\hat p_{\alpha} \bar
\partial \hat \theta^{\alpha} +\bar p^{\dot \alpha} \bar \partial
\bar \theta_{\dot \alpha}+\hat{\bar p}^{\dot \alpha} \bar \partial
\hat{\bar \theta}_{\dot \alpha} -\partial \rho \bar \partial \rho
\cr
+\lambda^a \partial \lambda^a+ \bar b\partial\bar c \Big]+
S_{\K3\times T^2}~.
\end{eqnarray}
Here $a=1,\dots, 28$ labels what is left of the 32 right-moving,
real, chiral fermions making up the root lattice of the heterotic
rank 16, 496-dimensional gauge group $G$ after we subtract 2 complex
dimensions for the right-moving $\K3$ fermions necessary for anomaly
cancelation.
The fermionic fields $(\bar b, \bar c)$ are the usual right-moving
conformal ghosts. The first line corresponds to usual Green-Schwarz
part. The second describes the 28
right-moving chiral fermions and the lagrangian for the $\K3\times T^2$. 

The construction of the $\K3 \times T^2$ $\sigma$-model proceeds
exactly as that for the Calabi-Yau \cite{Berkovits:1994wr}. The
super-$\K3$ coordinates are given by two $N=(2,0)$ worldsheet chiral
superfields $Y^i=y^i+\kappa^+ \Psi^i+\dots$ and $\bar Y^{\bar \imath}=\bar y^{\bar \imath}+ \kappa^-\bar\Psi^{\bar \imath}+\dots$ The coordinate of the super-torus will be
denoted by $X=x+\kappa^+ \psi+\dots$ and $\bar X=\bar x+ \kappa^-\bar\psi+\dots$ and the heterotic fermions $\Lambda^A=\lambda^A+\dots$ with $A=1,2$. The action is
\begin{eqnarray}
\label{Sigma} 
S_{\K3\times T^2}=\int {\rm d} z{\rm d}\bar z {\rm d}\kappa^+ {\rm d}\kappa^- \Big[ \partial_i K \bar \partial Y^i- \partial_{\bar \imath} K
\bar \partial \bar Y^{\bar \imath}+ \bar \Lambda^{\bar A}({\rm
e}^V)_{\bar A B} \Lambda^{B}
+\bar X \bar \partial X 
- X \bar \partial \bar X \Big].
\end{eqnarray}
where $K[Y, \bar Y]$ is the K\"ahler potential for $\K3$ and $V_{\bar A B}$ is the vector bundle background for the heterotic fermions. In
writing the $\sigma$-model (\ref{Sigma}) we have used the fact that
the torus is flat to set the $\U1$ part of the gauge connection to
zero.

The RNS fermion on the torus $\psi$ is bosonized as $\psi={\rm e}^\sigma$ and has
conformal weight $(\frac 12,0)$. The formalism requires a time-like
chiral boson $\rho$. The exponential ${\rm e}^\rho$ has conformal
weight $(-\frac 12,0)$.  The action excluding the $\K3$ factor is free and gives\footnote{As usual, the chirality of $\rho$ is understood to be imposed by hand.}
\begin{eqnarray}
x^m x^n\sim \eta^{mn} \log |z|^2~~~,~~~ p_{  \alpha} \theta^{  \beta}
\sim \frac 1z \delta_{  \alpha}^{  \beta}~~~,~~~ \rho \rho\sim -\log
z~~~,~~~\cr
\sigma \sigma\sim \log z~~~,~~~ x \bar x\sim \log |z|^2~~~,~~~\lambda^a
\lambda^b\sim \frac 1{\bar z} \delta^{ab} ~~~,
\end{eqnarray}
for the OPEs of the worldsheet fields.

The right-moving worldsheet stress-energy is
\begin{eqnarray}
\bar T= \bar \partial x_m \bar \partial x^m +\bar \partial x \bar \partial \bar x +\lambda_a \bar \partial
\lambda^a+ \bar c\bar\partial \bar b +2\bar\partial \bar c \bar b+ \bar T_{\K3}~.
\end{eqnarray}
On the other hand, the left-moving symmetry is the $N=(2,0)$ superconformal algebra and 
consists of the tensors $\{J, G^\pm, T\}$ which are required to
generate the familiar $N=2$ algebra, which we will now construct.
Following the establishment of the covariant 6-dimensional formalism
\cite{Berkovits:1999du}, we introduce the projective constraints
\begin{eqnarray}
\label{ProjectiveConstraints}
\nabla_\alpha=\zeta d_\alpha-\hat
d_\alpha=0~~~,~~~ \bar \nabla_{\dot \alpha}=-\frac 1\zeta  \bar d_{\dot
\alpha}- \hat{\bar d}_{\dot \alpha}=0~~~,~~~ \zeta={\rm
e}^{\rho-\sigma}~~~.
\end{eqnarray}
Note that the chiral boson $\rho$ and the bosonized RNS fermion 
are essential in the construction.
The covariant Siegel derivatives \cite{Siegel:1985xj} are shifted
compared to the hybrid formalism
 $d_\alpha\to d_\alpha +\frac 12 \hat \theta_\alpha \partial x+\dots$ to
include terms necessary to form the centrally extended superspace
algebra
\begin{eqnarray}
d_\alpha (z) \bar d_{\dot \alpha}(0) \sim {i\over z} \Pi_{\alpha \dot
\alpha} ~~~&,&~~~ d_\alpha (z) \hat d_\beta(0)\sim {1\over z}
\varepsilon_{\alpha \beta} {\bar \pi}~~~,\cr
\hat d_\alpha (z) \hat{\bar d}_{\dot \alpha}(0) \sim {i\over z}
\Pi_{\alpha \dot \alpha} ~~~&,&~~~ \bar d_{\dot \alpha} (z) \hat{\bar
d}_{\dot \beta}(0)\sim {1\over z} \varepsilon_{\dot \alpha \dot
\beta}\pi ~~~,\cr
d_\alpha(z) \Pi_{\beta \dot \beta}(0)\sim {i\over
z}\varepsilon_{\alpha \beta}
\partial \bar \theta_{\dot \beta} ~~~&,&~~~ d_{\alpha}(z) \pi\sim {1\over z}
\partial \hat \theta_\alpha ~~~,\cr
\hat d_{\alpha}(z) \Pi_{\beta \dot \beta}(0)\sim {i\over
z}\varepsilon_{\alpha \beta} \partial\hat {\bar\theta}_{\dot \beta}
~~~&,&~~~ \hat d_{\alpha}(z)\pi\sim - {1\over z} \partial
\theta_\alpha ~~~,\cr
\Pi^m (z) \Pi^n(0)\sim {1\over z^2} \eta^{mn}~~~&,&~~~ \bar \pi(z)
\pi(0)\sim {1\over z^2}~~~.
\end{eqnarray}
All other OPEs vanish or are related to these by conjugation. The
4-dimensional covariant momenta are defined as
\begin{eqnarray}
\Pi_{\alpha \dot \alpha}&=&\partial x_{\alpha \dot \alpha}+\frac i2
\left( \theta_\alpha \stackrel\leftrightarrow \partial \bar
\theta_{\dot \alpha}+ \hat \theta_\alpha \stackrel\leftrightarrow
\partial {\hat {\bar \theta}}_{\dot \alpha}\right)~,\cr 
{\bar \pi}& =&\partial x+\frac 12 \hat{\bar \theta}_{\dot \alpha}
\stackrel\leftrightarrow\partial \bar \theta^{\dot \alpha}~,\cr\cr 
{\pi}& =&\partial \bar x+\frac 12 \hat{\theta}^{\alpha}
\stackrel\leftrightarrow\partial  \theta_{\alpha}~.
\end{eqnarray}

The projective constraints (\ref{ProjectiveConstraints}) are required
to commute with the $N=2$ generators and can be used to gauge-fix
$(\hat \theta , {\hat {\bar \theta}}) \to 0$.  The $N=2$ constraints
should reduce in this gauge to the familiar covariant hybrid
constraints. Alternatively, setting $(\nabla, \bar \nabla)\to 0$
reduces the constraints to their $\O2$-symmetric form
\cite{Berkovits:1999du} c.f.~equation (\ref{o2}). The generators
satisfying these two requirements are given by\footnote{Another
useful form is
\begin{eqnarray}
\label{o2} T &=&\frac12 \Pi^{\alpha \dot \alpha} \Pi_{\alpha \dot
\alpha}+d^\alpha \partial \theta_\alpha+ \hat d^\alpha \partial \hat
\theta_\alpha  + {\bar  d}_{\dot \alpha}  \partial \bar \theta^{\dot
\alpha}+ \hat{\bar d}_{\dot \alpha} \partial \hat{\bar
\theta}{}^{\dot \alpha}-\frac12
\partial \rho \partial \rho\cr
&&+\bar \pi \pi +{1\over 2} \bar \psi
\stackrel{\leftrightarrow}{\partial}\psi +\nabla^\alpha \partial \hat
\theta_{\alpha} +\bar \nabla_{\dot \alpha}\partial {\hat {\bar
\theta}}{}^{\dot \alpha}+T_{\K3}~.
\end{eqnarray}
The limit $\nabla \to 0$ gives the $O(2)$ symmetric stress tensor. }
\begin{eqnarray}
\label{n2algebra} T &=&\frac12 \Pi^{\alpha \dot \alpha} \Pi_{\alpha
\dot \alpha}+d^\alpha \left( \partial \theta_\alpha+ \zeta  \partial
\hat \theta_\alpha \right) +{\bar  d}_{\dot \alpha} \left( \partial
\bar \theta^{\dot \alpha}-\frac1\zeta \partial \hat{\bar
\theta}^{\dot \alpha}\right) -\frac12 \partial \rho \partial \rho \cr
&&\hspace{2.6in}+\bar \pi \pi+ \frac 12 \partial \sigma \partial
\sigma +T_{\K3}~~~,\cr
G^+&=& {\rm e}^\rho d^2 +  {\rm e}^{\sigma} \bar \pi+ G^+_{\K3}
~~~,\cr
G^-&=& {\rm e}^{- \rho} \bar d^2 +{\rm e}^{- \sigma} \pi +G^-_{\K3}
~~~,\cr
J&=&-\partial( \rho-\sigma) +J_{\K3}~~~.
\end{eqnarray}
In the gauge $\hat \theta, {\hat {\bar \theta}}=0$ the
$N=2$ constraints (\ref{n2algebra}) reduce to the familiar
four-dimensional hybrid constraints which commute with the Siegel
supercharges extended by
\begin{eqnarray}
q_\alpha\to q_\alpha -\oint \hat \theta_\alpha \pi~~~&,&~~~ \bar
q_{\dot \alpha}\to\bar q_{\dot \alpha}-\oint {\hat {\bar
\theta}}_{\dot \alpha} \bar \pi~~~,\cr
\hat q_\alpha\to \hat q_\alpha -\oint \theta_\alpha \pi~~~&,&~~~
\hat{\bar q}_{\dot \alpha}\to\hat{\bar q}_{\dot \alpha}-\oint {{\bar
\theta}}_{\dot \alpha} \bar \pi~~~.
\end{eqnarray}
Note that
\begin{eqnarray}
T_\mathrm{comp}=T_{\K3}\,+ \bar \pi \pi+\frac 12 \bar \psi
\stackrel{\leftrightarrow}{\partial} \psi ~~~,~~~J_\mathrm{comp}=J_{\K3}\;+\psi \bar \psi ~~~,\cr
G^+_\mathrm{comp}=G^+_{\K3}+ \psi \bar \pi ~~~,~~~G^-_\mathrm{comp}=G^-_{\K3}+ \bar \psi \pi ~~~,~~~
\end{eqnarray}
forms an $N=(2,0)$ superconformal algebra with $c=9$.
In the 6-dimensional case \cite{Berkovits:1999du}, Lorentz invariance
requires that $G^+$ be of order 4 in $d$ while $G^-$ is of order 0.
There the fact that $G^+$ commutes with the projective constraint
was attributed to the fact that it could be shown to be itself
proportional to the fourth power of the constraint $G^+\propto
\nabla(\nabla(\nabla(\nabla({\rm e}^{-2\rho+3\sigma}))))$({note that $\nabla$, $\rho$ and $\sigma$ have different definitions in \cite{Berkovits:1999du}}). By
contrast, in the case of dimension 4, $d^2$ and $\bar d^2$ are
separately Lorentz invariant which allows the constraint algebra
(\ref{n2algebra}) to take a symmetric form. It should therefore not
be surprising that we find
\begin{eqnarray}
\label{NiceForm} G^+=-\frac12 \varepsilon^{\alpha \beta}
\nabla_\alpha \left(\nabla_\beta\left( {\rm e}^{-\rho+2\sigma}\right)
\right) ~,~ G^-=-\frac12 \varepsilon^{\dot \alpha \dot \beta} \bar
\nabla_{\dot \alpha} \left(\bar \nabla_{\dot \beta}\left( {\rm
e}^{\rho-2\sigma}\right) \right)~,
\end{eqnarray}
in analogy with the 6-dimensional case. This observation will be
useful in the construction of vertex operators in section
\ref{Section:Supermodulons}.

\section{Vertex Operators}
\label{Sec:VtexOps}
A general real (irreducible) integrated vertex operator $V$ in a
background of the product form $ M\times \K3$ factorizes in the
absence of flux as
\begin{eqnarray}\label{vertexone}
\int {\rm d}\mu \, \mathbf  U(x, \theta, \bar \theta, \hat \theta,
{\hat{\bar \theta}}; {\zeta}) \, \mathcal O ~.
\end{eqnarray}
Here ${\rm d}\mu$ denotes the world-sheet supermeasure, $\mathbf U$
is a 6-dimensional space-time superfield, that is, a function of
the $\{x^m,x, \bar x,\theta, \bar \theta, \hat \theta, {\hat{\bar
\theta}} \}$ 0-modes, and $\mathcal O=\mathcal O_L\otimes \mathcal
O_R$ is an operator in the conformal field theory. The vertex
operator (\ref{vertexone}) is required to satisfy the following conditions: It
\begin{enumerate}
\item has no poles with $J$, \label{C:J}
\item commutes with the projective constraints (\ref{ProjectiveConstraints}), \label{C:Projective}
\item is world-sheet supersymmetric, and \label{C:Susy}
\item has conformal weight $(0,0)$. \label{C:Conformal}
\end{enumerate}
Condition \ref{C:J} means that $\mathbf U$ depends on the chiral
scalars $\rho$ and $\sigma$ only in the combination ${\zeta}$ and is
therefore a general power series in the latter.
This condition,
together with requirement \ref{C:Projective} is the definition of
a general projective superfield \frenchspacing{c.f. sections}
\ref{Section:CIVtex} and \ref{Section:Supermodulons}.\footnote{In particular,
$\zeta$ plays the role of the projective superspace harmonic and is related
to the RNS torus fermion by $\zeta=\mathrm e^\rho\bar \psi$. In this sense,
the projective parameter has a geometric interpretation -- almost.}
Condition
\ref{C:Susy} can be met by writing either $\int {\rm d}\mu = \int
{\rm d}z {\rm d}\bar z{\rm d}\kappa^+ {\rm d}\kappa^-$ with a general
worldsheet superfield as integrand or $\int {\rm d}\mu = \int {\rm
d}z {\rm d}\bar z {\rm d}\kappa^+$ (or its conjugate) with a worldsheet
(anti-)chiral one. We then interpret $\int {\rm d}\kappa^+ \mathbf U
\mathcal O = G^+(\mathbf U \mathcal O)$, $\int {\rm d}\kappa^-
\mathbf U \mathcal O = G^-(\mathbf U \mathcal O)$, {\it et cetera}.
As usual, the worldsheet chiral measures require worldsheet chiral
integrands in order for the integrated vertex operators to be
supersymmetric while the integrand in the case of the full measure is
real but otherwise unconstrained. Finally, condition \ref{C:Conformal} is really a
condition on the operator $\mathcal O$ since $\mathbf U$ depends only
on ${\zeta}$ and the 0-modes of worldsheet fields.

Let us start by considering the full measure. Condition
\ref{C:Conformal} implies that $\mathcal O$ has conformal weight
$(0,1)$, implying that $\mathcal O_L=1$ and $\mathcal O_R\in \{\bar
\partial x_m,\bar \partial x,\bar\partial \bar x,  j^I\}$. Then the vertex operator reduces to $G^+\left(
G^-\left( \mathbf U \mathcal O\right)\right)= G^+\left( G^-\left(
\mathbf U \right)\right)\mathcal O$. The most general unconstrained
vertex operator is therefore of the form
\begin{eqnarray}
\label{E:CIVtex}
\int \mathrm d z\mathrm d\bar z \, \sum_{v} G^-\left( G^+\left(
\mathbf U^v \right)\right) \mathcal O_v 
\end{eqnarray}
where the sum runs over $v$ with $\mathcal O_m=\bar \partial x_m$,
$\mathcal O_I=j_I$, $\mathcal O_x=\bar\partial x$ and $\mathcal O_{\bar x}=\bar\partial \bar x$. These operators are independent of the
compactification manifold and we elaborate on their structure in
subsection \ref{Section:CIVtex}. The projective superfields ${\mathbf U}^v$ have arbitrary analytic dependence on $\zeta$ and represent the supergravity and the gauge part of spectrum. 
Note that we did not include operators of the form $\bar\partial \theta^\alpha$ since 
they vanish on shell.

The construction of (anti)chiral operators is analogous. In this case we have $\int {\rm d}\kappa^+
\mathbf \Phi \mathcal O = G^+(\mathbf \Phi \mathcal O)$ and the
conformal weight of $\mathcal O$ is $(\frac12,1)$. As above, we are assuming that $\mathbf\Phi$ depends on chargeless weight zero worldsheet fields $\{x^m,x, \bar x,\theta, \bar \theta, \hat \theta, {\hat{\bar
\theta}},\zeta \}$ and commutes with the projective constraints. In this case $\mathcal O$ has to have charge $-1$, 
which means that the left-moving part depends on $\bar \Psi^{\bar \imath}$. 
Worldsheet supersymmetry requires that $G^-
(\mathbf \Phi \mathcal O)=0$. In the spacetime part, by the relation (\ref{NiceForm})
between the projective constraint and the $G^-$ operator and the condition that $\mathbf \Phi$ commutes with (\ref{ProjectiveConstraints}), we find that $\mathbf \Phi$ must have no poles with ${\rm e}^{\rho-2\sigma}$.
This, in turn, implies that it has no negative powers of
${\zeta}$.\footnote{This is precisely the mechanism by
which the 6-dimensional type II hybrid string description gives rise
to chiral and twisted-chiral superfields \cite{Berkovits:1999du}.}
Such analytic projective superfields ${\mathbf\Phi}=\sum_{n\ge0}{\zeta}^n
\Phi_n$ are called ``arctic''. The compactification dependent
supermodulons are of this type and we will return to them in section
\ref{Section:Supermodulons}. 
For the operator $\mathcal O$, chirality implies that $G^-_{\K3}({\mathcal O})=0$ and, furthermore, should be in the cohomology 
of $G^-_{\K3}$ {to avoid pure gauge deformations}. The supermodulus vertex operator is therefore of the form
\begin{eqnarray}
\label{E:Supermodulon} 
\int {\rm d}z {\rm d}\bar z \, \sum_{\omega}
G^+\left( \mathbf \Phi^{\omega} \cal O_{\omega}
\right)
\end{eqnarray}
where the sum runs over the cohomology of $G^-_{\K3}$, $\mathcal O_\omega$ is the operator representing the cohomology class and $\mathbf \Phi^\omega$ is an arctic modulus field. 
We now turn to a more detailed description of this general setup.

\subsection{Compactification independent vertex operators}
\label{Section:CIVtex} As we have just derived, the compactification
independent vertex operators (\ref{E:CIVtex}) are all of the form
$G^-(G^+(\mathbf U))$ times a current $1\otimes \mathcal
O_R$. The superfield ${\bf U}(x, \theta, \bar \theta, \hat \theta,
\hat{\bar \theta}, {\zeta})$ satisfies the projective constraint
\begin{eqnarray}
\label{ProjectiveConstraintstraint} \nabla_{\alpha}
\mathbf{U}=0~~~,~~~ \bar \nabla_{\dot \alpha} \mathbf{U}=0~.
\end{eqnarray}
A superfield $\mathbf U$ holomorphic in an auxiliary variable
${\zeta}$ and satisfying this condition is called projective
\cite{Karlhede:1984vr}. Projective superfields can be expanded in
harmonics as
\begin{eqnarray}
\label{HarmExp} {\bf U}&=&\sum_{n=-\infty}^\infty {\zeta}^n U_n(x,
\theta, \bar \theta, \hat \theta, \hat{\bar \theta})
\end{eqnarray}
where the $U_n$ are ordinary $N=2$ superfields. The projective
superspace is an extension of $\O2$ superspace by an auxiliary
complex projective line ${\mathbb C}P^1$ parameterized (in the
northern patch) by ${\zeta}$. Superspace conjugation extends
naturally to projective superspace by the antipodal map on the
projective sphere. In coordinates it is defined to act as
${\zeta}\mapsto -1/{\zeta}$ and superspace conjugation on the rest.
If $\mathbf U$ is real with respect to this operation, this implies 
$U_{-n}=(-)^n\bar U_n$. Such a projective superfield is called {\em 
tropical}.

The projective constraints (\ref{ProjectiveConstraintstraint}) can be
used together with the harmonic expansion (\ref{HarmExp}) to hide the
dependence of the coefficient fields $U_n$ on $(\hat \theta,
\hat{\bar \theta})$ since they imply, for example, $\hat D_\alpha U_n
=D_\alpha U_{n+1}$. In this sense, we may think of projective
superfields as $N=1$ multiplets with `manifest' $N=2$ supersymmetry
and we will do so in what follows.

The operator $G^-(G^+(\mathbf U))$ is invariant under the linearized
transformation
\begin{eqnarray}
\label{GaugeTrans} \delta {\mathbf U}= {\mathbf
\Lambda}({\zeta})+\bar {\mathbf \Lambda}({\zeta})~,
\end{eqnarray}
where $\mathbf \Lambda$ is a projective superfield defined such that
it has no simple pole with the $G^-$ constraint. This implies in
particular that it does not depend on negative powers of ${\zeta}$.
The projective superfield ${\mathbf \Lambda}=\sum_{n=0}^\infty
{\zeta}^n \Lambda_n$ is called {\em arctic} and its projective conjugate is called antarctic. The
projective constraint implies that the lowest components $\Lambda_0$
and $\Lambda_1$ of the arctic superfield are constrained in the $N=1$
sense. Specifically,
\begin{eqnarray}
\label{N1Const} \bar D_{\dot \alpha} \Lambda_0=0~~~,~~~ \bar D^2
\Lambda_1=\partial \Lambda_0~,
\end{eqnarray}
in the presence of central charges $\partial$ (extra
dimensions).\footnote{\label{hyper}The second condition is equivalent
to the projective equation $\oint {{\rm d}\zeta \over \zeta}
\left(-\frac 1\zeta \bar D^2 +\partial\right) \mathbf \Lambda =0$.
This is the projective superspace analogue of the statement that
$\mathbf \Lambda$ has no simple pole with $G^-$. We should also
mention that starting with an arctic field $\mathbf \Lambda$ and
setting $\Lambda_n=0 \forall n\ge 2$, the second constraint
(\ref{N1Const}) is strengthened to $D_\alpha \Lambda_1=0$.} In $N=1$
notation, the gauge transformations take the expected form
\begin{eqnarray}
\label{N1GaugeTrans} \delta U_0= \Lambda_0+\bar \Lambda_0~~~,~~~
\delta U_1=\Lambda_1~.
\end{eqnarray}
We will return to multiplets of this type in subsection \ref{Section:Supermodulons}.

The integrated vertex operator $\int \mathrm dz\mathrm d\bar z
G^-(G^+(\mathbf U))$ can be expanded over worldsheet fields as
\cite{Siegel:1985xj}\footnote{
The
original proposal for the 6-dimensional open superstring vertex
operator \cite{Berkovits:1994vy,Berkovits:1999du} was of the same
form as the one proposed here (\ref{OpenVtex}). It was later argued
\cite{Berkovits:2000fe} that one must include terms of the form $\int
(u\gamma^{mn}v) F_{mn}$ where $u$ is the ghost for the projective
constraint and $v$ is its momentum. This was necessary because the
6-dimensional Lorentz generator is of the form $M^{mn}=\theta
\gamma^{mn} p+u \gamma^{mn} v$ where the ghost part is needed to
produce the correct numerical coefficient in the double pole of $MM$
(Lorentz invariance). It was also pointed out that this term is
unnecessary in 4 dimensions since there $\theta \gamma^{mn} p$
already produces the correct pole structure. The hybrid formalism for the
heterotic string presented here does not have manifest 6-dimensional
Lorentz symmetry and therefore does not require the ghost correction
to the Lorentz generator or the open string vertex operator. }
\begin{eqnarray}
\label{OpenVtex} V&=&\int {\rm d}z {\rm d}\bar z \ \Big( \Pi^m
A_m+\pi A+\bar \pi \bar A +\partial \theta^\alpha
\Gamma_\alpha+\partial \bar \theta_{\dot \alpha} \bar \Gamma^{\dot
\alpha} \cr&&\hspace{1.2in} +d_\alpha W^\alpha+\bar d^{\dot
\alpha}\bar W_{\dot \alpha} +{\zeta} d_\alpha Z^\alpha -\frac
1{\zeta} \bar d^{\dot \alpha} \bar Z_{\dot \alpha} \Big)~,
\end{eqnarray}
with gauge covariant potentials
\begin{eqnarray}
A_m= (\sigma_m)^{\alpha \dot \alpha} \left[ D_\alpha , \bar D_{\dot
\alpha}\right] U_0 ~,~A=\bar D^2 U_1+ \partial
U_0~,~\Gamma_\alpha=D_\alpha U_0~,
\end{eqnarray}
and field strengths
\begin{eqnarray}
W_\alpha= \bar D^2 D_\alpha U_0~~~,~~~Z_\alpha=D_\alpha A~.
\end{eqnarray}
Note that the $N=1$ constraints (\ref{N1Const}) and gauge
transformations (\ref{N1GaugeTrans}) together imply that $W_\alpha$
and $Z_\alpha$ are invariant and that $\delta A_m=\partial_m \alpha$
and $\delta A=\partial \alpha$ for $\alpha$ proportional to the
imaginary part of the lowest component of $\Lambda_0$.

The ``left-moving'' vertex operator (\ref{HarmExp}) needs to be
completed by the right-moving currents $\bar \partial x_m$, $\bar
\partial x$, $\bar \partial\bar x$ and the gauge currents $j^I$.
As usual in the hybrid formalism, unintegrated vertex operators
correspond to superspace prepotentials yielding, respectively, the
conformal supergravity prepotential $\mathbf U^m$, a complex vector
prepotential $\mathbf A+i\mathbf B$ coming from the torus, and the super-Yang-Mills
prepotential $\{\mathbf V^I\}_{I=1}^{381}$ all with $N=(1,0)$ supersymmetry in 6
dimensions.

\subsection{Compactification dependent vertex operators}
\label{Section:Supermodulons}
The second type of vertex operator is chiral on the worldsheet. The
integrated form of this operator (\ref{E:Supermodulon}) uses an
arctic superfield $\mathbf \Phi=\sum_{n=0}^\infty {\zeta}^n \Phi_n$
with the lowest components $\Phi_0$ and $\Phi_1$ obeying $N=1$-type
constraints (\ref{N1Const}). This multiplet is the projective
superspace version of a hypermultiplet. We pause here to elaborate a
bit on this point.

What is traditionally referred to as a hypermultiplet $H$ decomposes
under $N=2\to 1$ into a chiral field $q^1$ and an antichiral field
$q^2$ forming a doublet under the $\SU2$ $R$-symmetry. In
4-dimensional, $N=1$ superspace a chiral field $\Phi$ is Poincar\'e
dual to a complex linear field (also known as the non-minimal scalar
multiplet) $\bar \Gamma$ in the sense that the constraint of $\Phi$
is the equation of motion of $\bar \Gamma$ and {\it vice versa}.
Indeed, complex linear fields obey $\bar D^2 \Gamma=0$ off-shell,
which would have been the equation of motion had $\Gamma$ been a
free chiral field. In the presence of central charge, this constraint is
modified to $\bar D^2 \Gamma=
\partial  \Phi$ which exactly reproduces the constraints (\ref{N1Const}) on the
first two components of an arctic field. The precise statement is,
therefore, that an arctic field $\mathbf \Phi$ is the off-shell
extension of a half-dualized hypermultiplet. Alternatively, a
hypermultiplet {\em is} an on-shell arctic field in which all the
auxiliary components have been integrated out (see footnote
\ref{hyper}).

We see, then, that in this formulation the supermoduli do not appear
naturally as hypermultiplets but as off-shell extensions of
chiral-non-minimal multiplets \cite{Gates:1998si}. In order to use the power of the
quaternionic-K\"ahler structure of the hypermultiplet moduli space,
we must first integrate out the infinite number of auxiliary
superfields which puts some of the supersymmetry on-shell and then
perform a duality transformation.

In the large radius limit, $G^-_{\K3} \sim d\bar y^{\bar \imath}\partial / \partial \bar y^{\bar \imath}$ becomes the Dolbeault operator so the compactification dependent spectrum consists of artic superfields 
coupling to operators in  the cohomology of $(0,1)$-forms $\omega_{\bar \imath}^\bullet$ taking values in various vector bundles over $\K3$ \cite{Distler:1987ee}. The choice of the tangent space corresponds to the choice of right moving part of the operator $\mathcal O$. There are only three choices that give a non-empty cohomology. First there is $\omega_{\bar \imath}^{\bar k}$ coupling to $\bar\partial y^k$, which gives the 
20 artic modulus multiplets. We also have deformations of the vector bundle, described by $\omega_{\bar \imath}^{A\bar B}$, coupling to 
$\lambda^A \bar\lambda^{\bar B}$ which gives 45 artic multiplets. Finally we have $\omega_{\bar \imath}^{(A, s)}$, where $(A,s)$ 
denotes the $(\mathbf 2,\mathbf{56})$-dimensional representation of $SU(2)\times E_7$, coupling to 
currents $j_{A,s}$ constructed from 16 of the original 32 heterotic fermions. 
Putting all of this together, we obtain a more explicit form for the vertex operator (\ref{E:Supermodulon}) 
\begin{eqnarray}
V=\int {\rm d}z {\rm d}\bar z \, \sum_{\omega^\bullet}
G^+\left( \mathbf \Phi^{\omega^\bullet} \Omega_{\omega^\bullet}(y)
\right)~,
\end{eqnarray}
where $\Omega_{\omega^\bullet}$ denotes the operator corresponding to the representative $\omega^\bullet$ of the cohomology group and $\mathbf \Phi^{\omega^\bullet}$ is the spacetime field dual to it. 
Of course, all of these operators come with their complex conjugates.

\paragraph{Spectrum}
Let us summarize the spectrum of the heterotic $\K3\times T^2$ vacuum as it follows from the analysis above. The complete list of tropical fields is 
\begin{eqnarray}
\label{Eq:Tropical}
\mathbf U^m \oplus \mathbf A\oplus \mathbf B\oplus \{\mathbf V^I\}_{I=1}^{381}
\end{eqnarray}
with the index $I$ running over the adjoint representation of $E_7\times E_8$. The gravitational prepotential contains two $\U 1$ gauge fields, one of which is the graviphoton and the other sits in a vector-tensor multiplet \cite{deWit:1995zg} which contains the $B$-field and the dilaton $\phi$. After dualizing the $B$-field into an axion $a$ these scalars are often grouped into a complex scalar $S\sim \mathrm e^\phi+i a$. Two more gauge fields sit in the complex field $\mathbf A+i\mathbf B$ which also contains four real scalars, the vacuum expectation values of which parameterize the $T^2$ moduli. They are usually grouped into the complex fields $T\sim \sqrt \gamma+i \beta$ and $U\sim (\sqrt \gamma-i \gamma_{12})/\gamma_{11}$ where $\gamma$ and $\beta$ denote the metric and $B$-field on the torus. Overall the gauge group is therefore $E_8\times E_7\times {\U 1}^4$.

The supermodulon spectrum is given in terms of arctic fields as
\begin{eqnarray}
\label{Eq:Arctic}
\{\mathbf \Upsilon_n\}_{n=1}^{20}\oplus \{\mathbf \Phi^\mathbf{1}_n\}_{n=1}^{45} \oplus\{ \mathbf \Phi_n^\mathbf{56}\}_{n=1}^{10}
\end{eqnarray}
Here the $\mathbf \Upsilon$s are the gauge group singlets coming from the moduli of the $\K3$. They couple to the forms in $H^{1,0}_{\bar \partial}(T\! S)\cong H^{1,1}_{\bar \partial}(S)$ and there are 20 such forms.
The $\mathbf \Phi$s come from the gauge bundle moduli. With the standard embedding $E_8\supset SU(2)\times E_7$ the adjoint representation decomposes as $\mathbf {248}\to \mathbf{(1,133)\oplus (3, 1)\oplus (2,56)}$. The first term is carried by $\mathbf V^I$ above. The second is a singlet under $E_7$, valued in the adjoint representation of $SU(2)$. The arctic field $\mathbf \Phi^\mathbf{1}$ couples to the forms in $H^{1,0}_{\bar \partial}(\mathrm{End}T\!S)$, the dimension of which is 45. Finally, there are 10 (quaternionic) moduli carried by $\mathbf \Phi^\mathbf{56}$ valued in the 56-dimensional representation of $E_7$.

Equations (\ref{Eq:Tropical}) and (\ref{Eq:Arctic}) constitute the spectrum of the heterotic $\K3\times T^2$ vacuum written in a compact and manifestly symmetric form {\it via} projective superspace. It is easy to see that the known component spectrum \cite{deWit:1995zg} is reproduced exactly. 

\section{Supergravity}
\label{Section:SG}
The 4-dimensional, $N=2$ gravitational multiplet $\mathbf U^m$
deserves some elaboration. The existence of this theory was predicted
by Siegel in \cite{Siegel:1995px} where it is shown that the
gravitational prepotential $\mathbf U^m$ with gauge
transformation\footnote{The second term, relative to those in equation (\ref{GaugeTrans}), comes from the right-moving
part.} $\delta \mathbf U^m=\mathbf \Lambda^m +\bar \mathbf
\Lambda^m+\partial^m \mathbf L$ represents the irreducible conformal
supergravity multiplet off-shell. (Actually, Siegel presents this
formalism in harmonic superspace which needs to be converted to
projective language using Kuzenko's method \cite{Kuzenko:1998xm}.)
On-shell, this multiplet becomes reducible, factorizing into a supergravity
multiplet with bosonic components $(h_{mn}, A_m)$ and a vector-tensor multiplet $(b_{mn}, A^\prime_m, \phi)$. This theory is non-standard in the sense that usually the conformal supergravity multiplet is reducible off-shell but irreducible on-shell.

In order to write down the supergravity action and coupling to matter we have to know the compensator superfield appropriate for this description. In
\cite{Siegel:1995px}, Siegel 
argued, based on superspace methods, that there is only one compensator, that it is
given by the sum of an arctic field $\mathbf \Sigma$ and its conjugate $\mathbf G=\mathbf \Sigma+\bar \mathbf \Sigma$, and that the low-energy supergravity action in the absence
of matter is given by 
\begin{eqnarray}
S=\oint {\mathrm d \zeta\over \zeta} \int \mathrm d^6x \mathrm
d^4\theta E^{-1} \mathbf G^2
\end{eqnarray}
with $E=\mathrm {sdet} \left(E_A{}^M\right)$ the super-viel-bein determinant.

From the worldsheet point of view, the compensator is the space-time field that couples to the worldsheet curvature. Since in the present case we have $N=(2,0)$ 
supersymmetry, the curvature is described by a chiral worldsheet superfield
${\mathbf r}_+$ and its complex conjugate $\mathbf{r}_-$ {defined from worldsheet covariant derivatives $(D, \bar D, D_+, D_-)$ as 
$[D,D_+]={\mathbf r}_+( M + iJ)$ and $[D,D_-]={\mathbf r}_-(M-iJ)$, where $(M,J)$ are Lorenz and $U(1)$ generators on the worldsheet}. 
{The coupling to the compensator is given by}
\begin{eqnarray}
\int d^2z \left(G^+ ({\mathbf r}_+ {\mathbf \Sigma})+
G^- ({\mathbf r}_- \bar{\mathbf \Sigma})\right)~.
\end{eqnarray}
By the same analysis as in the vertex operator discussion above, these worldsheet fields can only couple in a consistent way to arctic and antarctic spacetime fields.

The supergeometrical origin of this theory is remarkable. Supergravitational theories can be constructed from any superfield representation of the super-Poincar\'e group as follows \cite{Siegel:1978mj}: Pick a superfield $\varphi$ which contains a scalar at the component level.
Construct an action from this field which is invariant under rigid scale transformations but which has the wrong sign kinetic term. Finally, gauge the rigid scale transformations. This will be a theory of supergravity with compensator $\varphi$. In the case of old-minimal supergravity, the conformal compensator is a $4|1$-dimensional chiral field. However, there is an extremely simple way to lift such theories to $4|2$ (or $6|1$) dimensions \cite{Gates:1998si}: one simply replaces the chiral field $\Phi\to \mathbf \Phi(\zeta)=\Phi+\zeta \Gamma+\dots$ with an arctic superfield. It therefore appears that the formulation of $N=2$ supergravity produced by the heterotic superstring on $\K3\times T^2$ is the na\"ive projective extension of old-minimal supergravity. Unfortunately, as of this writing, the full supergravity theory has not been worked out. 
Nevertheless, knowing the compensator structure and the supergravity prepotential, we can guess much of the structure from its $N=1$ old-minimal analogue.

The vector multiplets coming from the coupling to the right-moving currents can be
included by the anomaly cancellation mechanism: We replace in the action \cite{Siegel:1995px}
\begin{eqnarray}
\mathbf G\to\tilde \mathbf G=\mathbf G
+ c_\mathrm{L} \mathbf \Omega_{\mathrm{L}}
+ c_\mathrm{YM} \mathbf \Omega_{\mathrm{YM}}
\end{eqnarray}
where the $\mathbf \Omega_\rho$ are Chern-Simons 3-forms for super-Lorentz ($\rho=\mathrm L$) and the super-Yang-Mills connections ($\rho=\mathrm{YM}$). They are defined to satisfy that the projective integrals $\oint {{\rm d}\zeta
\over \zeta} \left(-\frac 1\zeta \bar D^2 +\partial\right) \mathbf
\Omega_{\rho}=A_\rho$ are the gravitational/gauge anomaly 4-forms. Explicitly $A_\mathrm L= W_{\alpha \beta}W^{\alpha \beta}$ is the square of the Weyl tensor and $A_\mathrm {YM}= \sum_{I=1}^{381} W^IW^I$ is the square of the Yang-Mills field strength.
The constants $c_{\rho}$ are determined by the compactification and gauge bundles. 

The gravitational couplings of the supermoduli can proceed as usual. Since the moduli space of hypermultiplets is hyper-K\"ahler, its structure is determined by a holomorphic prepotential $\mathcal F(\mathbf \Phi)$ homogeneous of degree 2. The general form of the coupling to hypermultiplets is therefore (see e.g. \cite{Kuzenko:2005sz})
\begin{eqnarray}
\oint {\mathrm d \zeta\over \zeta} \int \mathrm d^6 x \mathrm d^4 \theta \mathbf \Sigma \mathcal F(\mathbf \Phi)~.
\end{eqnarray}

\section{Outlook}
\label{Sec:Outlook}

In this article we have put advanced a covariant description of the heterotic string with manifest $4|2$-dimensional space-time supersymmetry. We have attempted to show that such a description succinctly represents compactifications of the heterotic string on manifolds of the form $\K3\times T^2$ by considering explicitly the case of the $E_8\oplus E_8$ gauge algebra.
In the process, we have found that the $N=(2,0)$ worldsheet supersymmetry naturally favors projective superspace targets. The resulting description of supergravity is non-standard and, unfortunately, has not been worked out in any detail as of this writing. Further work in this direction has the potential to shed light on the relation to recent advances in our understanding of the type II hypermultiplet moduli space \cite{Vandoren} to which it is related by string-string duality. In the latter case, the analogue of the special geometry prepotential
is formulated in projective superspace in terms of tensor multiplets which arise naturally in the construction from the c-map (T-duality). By contrast, the projective superspace description of the heterotic string never mentions the tensor multiplet representation, favoring instead the arctic description of its dual, the hypermultiplet.

Hybrid formulations of superstring theories of the type presented here have become somewhat numerous over the past 10 years. In each case (as here) a formulation has been developed to address various questions in a specific vacuum. Although these hybrid strings have been checked to varying degrees to be related to the universal RNS formulation by field redefinitions, it is has become clear that they should be related to one another in a more direct way. Understanding these relations is tantamount to having a more explicit realization of the dualities between them. For example, relating the formalism presented here to the dual type II hybrid description of the superstring on Calabi-Yau 3-folds should involve explicitly the famous relation between holomorphic vector bundles over $\K3$ surfaces and $\K3$- and elliptically-fibred Calabi-Yau 3-folds. Eventually, one would like to understand all these low-dimensional hybrid formulations in terms of a covariant 10-dimensional formalism -- perhaps the (non-minimal) pure-spinor superstring. The formulation presented here, although far from connecting to a covariant 10-dimensional string, sits naturally between the 6-dimensional type II string with 16 supercharges manifest and the 8-supercharge type II description.

Besides the technical problem of understanding the relation between the various hybrid strings, there seem to be quite a few applications/extensions of this approach. One such application is the possibility, alluded to already in Siegel's work \cite{Siegel:1995px}, of tensoring together two strings of the $N=(2,0)$ type to obtain an $N=(2,2)$ string with 16 supercharges, 
which will be described by a combination of two projective parameters.
Although the resulting formulation is likely to give a partially on-shell description of superspace supergravity rather than the elusive off-shell realization, it may nevertheless be a useful tool in the construction of the latter. Irrespective of this hopeful attitude toward the construction of off-shell 16-supercharge superspaces, such a formulation should relate to the existing partially on-shell hybrid description of the 6-dimensional type II string on a $\K3$ surface.

A second application relates to superstring field theory. One of the achievements of the hybrid formalism has been the construction of a WZNW-like open superstring field theory \cite{SuperSFT} and, more recently, a Chern-Simons-like $N=1$ heterotic string string field theory has been put forward \cite{HetSFT}. {One can try to use the open string version of the present paper in the open string field theory formulation of \cite{SuperSFT}} to give a new description of non-abelian $N=2$ gauge theories.
A more ambituous goal  
is to use  the explicit realization (\ref{n2algebra}) of the left-moving $N=2$ algebra to plug into the heterotic string field theory of Berkovits, Okawa, and Zweibach \cite{HetSFT} to yield the first closed string field theory with $N=2$ supersymmetry in the target space.
The natural string field would then be a tropical superfield. Again unfortunately, the projective formalism has at present not been sufficiently developed to simply write down the string field theory and some guess work has to be done as the unconstrained prepotential formalism for Yang-Mills theory is not known in projective superspace even though it has been known in $4|2$-dimensional harmonic superspace for over 20 years \cite{Galperin:1984av}. 

Finally, a more concrete application which uses only on-shell information is the computation of amplitudes using the present description. One important ingredient is the CFT 0-mode measure.
The subtle point here is the left-moving measure. Since the new superspace
coordinates $(\hat \theta,\hat{\bar\theta})$ are not related to the
underlying RNS formalism it is natural to assume that they do not
participate in a physical amplitude. Furthermore, given the
projective superspace description presented above, spacetime $N=2$ supersymmetry will
be realized through the projective superspace integral. 
We are thus led to propose that the measure for the left-moving 0-modes is given by
\begin{eqnarray}
\langle \theta^2\bar\theta^2 j^{++} \zeta^{-1} c\partial c\partial^2 c
\rangle=1
\end{eqnarray}
where $j^{++}$ is an $\mathfrak {su}(2)$ current in the $N=4$ topological algebra of the $\K3$. We see once again that $\zeta$ appears as the projective parameter.

\section*{Acknowledgments}\

We thank Warren Siegel for many helpful discussions on the structure of the various $N=2$ supergravity theories. {We also thank S. James Gates, Jr. and Sergei Kuzenko for correspondence and comments}. The work of WDL3 is
supported by NSF grant numbers PHY 0354776 and DMS 0502267. BCV would like to thank the NSF for support (under grant number PHY0354776) during the time when much of this work was undertaken.


\end{document}